\documentclass{sigma}

\begin{document}

\renewcommand{\PaperNumber}{??}

\FirstPageHeading

\ShortArticleName{Poincare Algebra Extension}

\ArticleName{Poincare Algebra Extension with
Tensor Generator}



\Author{Dmitrij V. SOROKA and Vyacheslav A. SOROKA}
\AuthorNameForHeading{D.V. Soroka and V.A. Soroka}
\Address{Kharkov Institute of Physics and Technology, 61108 Kharkov, Ukraine}
\Email{dsoroka@kipt.kharkov.ua, vsoroka@kipt.kharkov.ua}


\Abstract{A tensor extension of the Poincar\'e algebra is proposed for the 
arbitrary
dimensions. Casimir operators of the extension are constructed. A possible
supersymmetric generalization of this extension is also found in the
dimensions $D=2,3,4$.}


\Classification{20-xx} 

\section{Introduction}

There are many examples for the tensor 'central' extensions of the
super-Poincar\'e algebra (see, for
example,~\cite{dafr,holpr,z,hugpol,vst,agit,ds,caip}). However, there also
exists
the tensor extension of the Poincar\'e algebra itself. In the present report we
give the example of such an extension with the help of the second rank tensor
generator (see also~\cite{gios1,gios2}). Such an extension makes common sense, 
since it
is homomorphic to the Poincar\'e algebra. Moreover, the contraction of the
extended algebra leads also to the Poincar\'e algebra. It is interesting
enough that the momentum square Casimir operator for the Poincar\'e algebra
under this extension ceases to be the Casimir operator and it is generalized
by adding the term containing linearly the angular momentum\footnote{Note that
this reminds the relation for the Regge trajectory, which connects the mass
square with the angular momentum.}. Due to this fact, an irreducible
representation of the extended algebra\footnote{Concerning the irreducible
unitary representations of the extended Poincar\'e group in $(1+1)$ dimensions
see, for example,~\cite{mr}.} has to contain the fields of the different
masses. This extension with non-commuting momenta has also something in common
with the ideas of the papers~\cite{sn,ya,hl} and with the non-commutative
geometry idea~\cite{c}.
It is also shown that for the dimensions $D=2,3,4$ the extended
Poincar\'e algebra allows a supersymmetric generalization.

\section{Extension of the Poincar\'e algebra}

The Poincar\'e algebra for the components of the rotations $M_{ab}$ and
translations $P_a$ in $D$ dimensions
\begin{eqnarray}
[M_{ab},M_{cd}]=(g_{ad}M_{bc}+g_{bc}M_{ad})-(c\leftrightarrow d),\nonumber
\end{eqnarray}
\begin{eqnarray}
[M_{ab},P_c]=g_{bc}P_a-g_{ac}P_b,\nonumber
\end{eqnarray}
\begin{eqnarray}\label{2.1}
[P_a,P_b]=0
\end{eqnarray}
can be extended with the help of the tensor 'central' generator $Z_{ab}$
in the following way:
\begin{eqnarray}
[M_{ab},M_{cd}]=(g_{ad}M_{bc}+g_{bc}M_{ad})-(c\leftrightarrow d),\nonumber
\end{eqnarray}
\begin{eqnarray}
[M_{ab},P_c]=g_{bc}P_a-g_{ac}P_b,\nonumber
\end{eqnarray}
\begin{eqnarray}
[P_a,P_b]=cZ_{ab},\nonumber
\end{eqnarray}
\begin{eqnarray}
[M_{ab},Z_{cd}]=(g_{ad}Z_{bc}+g_{bc}Z_{ad})-(c\leftrightarrow d),\nonumber
\end{eqnarray}
\begin{eqnarray}
[P_a,Z_{bc}]=0,\nonumber
\end{eqnarray}
\begin{eqnarray}\label{2.2}
[Z_{ab},Z_{cd}]=0,
\end{eqnarray}
where $c$ is some constant.
By taking a set of the generators $Z_{ab}$ as a homomorphism kernel, we obtain
that the extended Poincar\'e algebra (\ref{2.2}) is homomorphic to the usual
Poincar\'e algebra (\ref{2.1}). Moreover, in the limit ${c\to 0}$ the algebra
(\ref{2.2}) goes to the semi-direct sum of the commutative ideal $Z_{ab}$ and
Poincar\'e algebra (\ref{2.1}).

Casimir operators of the extended Poincar\'e algebra are
\begin{eqnarray}\label{2.3}
Z_{a_1a_2}Z^{a_2a_3}\cdots Z_{a_{2k-1}a_{2k}}Z^{a_{2k}a_1},\quad(k=1,2,\ldots);
\end{eqnarray}
\begin{eqnarray}\label{2.4}
P^{a_1}Z_{a_1a_2}Z^{a_2a_3}&\cdots&
Z_{a_{2k-1}a_{2k}}Z^{a_{2k}a_{2k+1}}P_{a_{2k+1}}\cr\nonumber\\
+cZ^{aa_1}Z_{a_1a_2}&\cdots&
Z_{a_{2k-1}a_{2k}}Z^{a_{2k}a_{2k+1}}M_{a_{2k+1}a},\quad(k=0,1,2,\ldots);
\end{eqnarray}
\begin{eqnarray}\label{2.5}
\epsilon^{a_1a_2\ldots a_{2k-1}a_{2k}}Z_{a_1a_2}\cdots Z_{a_{2k-1}a_{2k}},
\quad2k=D,
\end{eqnarray}
where $\epsilon^{a_1\ldots a_{2k}}$, $\epsilon^{01\ldots{2k-1}}=1$ is the 
totally
antisymmetric Levi-Civita tensor in the even dimensions $D=2k$.
In particular, there is a Casimir operator generalizing the momentum
square
\begin{eqnarray}\label{2.6}
P^aP_a+cZ^{ab}M_{ba},
\end{eqnarray}
which indicates that an irreducible representation of the extended algebra
contains the fields having the different masses. Note that for the extended
algebra there is no generalization of the Pauli-Lubanski vector of the
Poincar\'e algebra. The expressions (\ref{2.3}) and (\ref{2.4}) for the Casimir
operators are valid for the extended Poincar\'e algebra (\ref{2.2}) in the
arbitrary dimensions $D$, but the expression (\ref{2.5}) is only true for
the even dimensions $D=2k$.

Note that in the case of the extended two-dimensional Poincar\'e algebra the
Casimir operators (\ref{2.3}) and (\ref{2.4}) can be expressed
\begin{eqnarray}
Z_{a_1a_2}Z^{a_2a_3}\cdots Z_{a_{2k-1}a_{2k}}Z^{a_{2k}a_1}=2Z^{2k},\nonumber
\end{eqnarray}
\begin{eqnarray}
P^{a_1}Z_{a_1a_2}Z^{a_2a_3}&\cdots&
Z_{a_{2k-1}a_{2k}}Z^{a_{2k}a_{2k+1}}P_{a_{2k+1}}\cr\nonumber\\
+cZ^{aa_1}Z_{a_1a_2}&\cdots&
Z_{a_{2k-1}a_{2k}}Z^{a_{2k}a_{2k+1}}M_{a_{2k+1}a}=Z^{2k}(P^aP_a+cZ^{ab}M_{ba})
\nonumber
\end{eqnarray}
as degrees of the following generating Casimir operators:
\begin{eqnarray}
Z={1\over2}\epsilon^{ab}Z_{ab},\nonumber
\end{eqnarray}
\begin{eqnarray}
P^aP_a+cZ^{ab}M_{ba},\nonumber
\end{eqnarray}
where $\epsilon^{ab}=-\epsilon^{ba}$, $\epsilon^{01}=1$ is the completely 
antisymmetric
two-dimensional Levi-Civita tensor. In the case of the extended
three-dimensional Poincar\'e algebra these Casimir operators can be expressed
\begin{eqnarray}
Z_{a_1a_2}Z^{a_2a_3}\cdots Z_{a_{2k-1}a_{2k}}Z^{a_{2k}a_1}=2(Z^aZ_a)^k,
\nonumber
\end{eqnarray}
\begin{eqnarray}
P^{a_1}Z_{a_1a_2}Z^{a_2a_3}&\cdots&
Z_{a_{2k-1}a_{2k}}Z^{a_{2k}a_{2k+1}}P_{a_{2k+1}}\cr\nonumber\\
+cZ^{aa_1}Z_{a_1a_2}&\cdots&
Z_{a_{2k-1}a_{2k}}Z^{a_{2k}a_{2k+1}}M_{a_{2k+1}a}\cr\nonumber\\
&=&(Z^aZ_a)^k(P^aP_a+cZ^{ab}M_{ba})-(Z^aZ_a)^{k-1}(P^aZ_a)^2
\nonumber
\end{eqnarray}
in terms of the following generating Casimir operators:
\begin{eqnarray}
Z^aZ_a,\nonumber
\end{eqnarray}
\begin{eqnarray}
P^aP_a+cZ^{ab}M_{ba},\nonumber
\end{eqnarray}
\begin{eqnarray}
P^aZ_a,\nonumber
\end{eqnarray}
where
\begin{eqnarray}
Z^a={1\over2}\epsilon^{abc}Z_{bc}\nonumber
\end{eqnarray}
and $\epsilon^{abc}$, $\epsilon^{012}=1$ is the totally antisymmetric 
three-dimensional
Levi-Civita tensor.
In the case of the extended $D$-dimensional ($D\ge4$) Poincar\'e algebra
the Casimir operators (\ref{2.3}) and (\ref{2.4}) can not be expressed in
terms of the finite number of the generating Casimir operators.

Generators of the left shifts, acting on the function $f(y)$ with a group
element $G$,
\begin{eqnarray}
[T(G)f](y)=f(G^{-1}y),\quad y=(x^a,z^{ab})\nonumber
\end{eqnarray}
have the form
\begin{eqnarray}
P_a=-\left({\partial\over\partial x^a}+
{c\over2}x^b{\partial\over\partial z^{ab}}\right),\nonumber
\end{eqnarray}
\begin{eqnarray}
Z_{ab}=-{\partial\over\partial z^{ab}},\nonumber
\end{eqnarray}
\begin{eqnarray}\label{2.7}
M_{ab}=x_a{\partial\over\partial x^b}-x_b{\partial\over\partial x^a}
+{z_a}^c{\partial\over\partial z^{bc}}-{z_b}^c{\partial\over\partial z^{ac}}
+S_{ab},
\end{eqnarray}
where coordinates $x^a$ correspond to the translation generators $P_a$,
coordinates $z^{ab}$ correspond to the generators $Z_{ab}$ and
$S_{ab}$ is a spin operator.

On the other hand, generators of the right shifts
\begin{eqnarray}
[T(G)f](y)=f(yG)\nonumber
\end{eqnarray}
have the form
\begin{eqnarray}
D_a\mathrel{\mathop=^{\rm def}}
{P_a}^r={\partial\over\partial x^a}-
{c\over2}x^b{\partial\over\partial z^{ab}},\nonumber
\end{eqnarray}
\begin{eqnarray}\label{2.8}
{Z_{ab}}^r=-Z_{ab}={\partial\over\partial z^{ab}}.
\end{eqnarray}
By taking into account (\ref{2.7}) and (\ref{2.8}), the Casimir operators
(\ref{2.4}) can be rewritten with the help of the generators $D_a$ in the
following way:
\begin{eqnarray}
D^{a_1}Z_{a_1a_2}Z^{a_2a_3}&\cdots&
Z_{a_{2k-1}a_{2k}}Z^{a_{2k}a_{2k+1}}D_{a_{2k+1}}\cr\nonumber\\
+cZ^{aa_1}Z_{a_1a_2}&\cdots&
Z_{a_{2k-1}a_{2k}}Z^{a_{2k}a_{2k+1}}S_{a_{2k+1}a},\quad(k=0,1,2,\ldots).
\nonumber
\end{eqnarray}

Note that the algebra
\begin{eqnarray}
[M_{ab},M_{cd}]=(g_{ad}M_{bc}+g_{bc}M_{ad})-(c\leftrightarrow d),\nonumber
\end{eqnarray}
\begin{eqnarray}
[M_{ab},P_c]=g_{bc}P_a-g_{ac}P_b,\nonumber
\end{eqnarray}
\begin{eqnarray}
[M_{ab},D_c]=g_{bc}D_a-g_{ac}D_b,\nonumber
\end{eqnarray}
\begin{eqnarray}
[P_a,P_b]=cZ_{ab},\nonumber
\end{eqnarray}
\begin{eqnarray}
[D_a,D_b]=-cZ_{ab},\nonumber
\end{eqnarray}
\begin{eqnarray}
[P_a,D_b]=0,\nonumber
\end{eqnarray}
\begin{eqnarray}
[M_{ab},Z_{cd}]=(g_{ad}Z_{bc}+g_{bc}Z_{ad})-(c\leftrightarrow d),\nonumber
\end{eqnarray}
\begin{eqnarray}
[P_a,Z_{bc}]=0,\nonumber
\end{eqnarray}
\begin{eqnarray}
[D_a,Z_{bc}]=0,\nonumber
\end{eqnarray}
\begin{eqnarray}\label{2.9}
[Z_{ab},Z_{cd}]=0,
\end{eqnarray}
formed by the generators $M_{ab}$, $P_a$, $D_a$ and $Z_{ab}$, has as
Casimir operators the operators (\ref{2.3}) and the following operators:
\begin{eqnarray}\label{2.10}
(P-D)^{a_1}Z_{a_1a_2}Z^{a_2a_3}&\cdots&
Z_{a_{2k-1}a_{2k}}Z^{a_{2k}a_{2k+1}}(P+D)_{a_{2k+1}}\cr\nonumber\\
+cZ^{aa_1}Z_{a_1a_2}&\cdots&
Z_{a_{2k-1}a_{2k}}Z^{a_{2k}a_{2k+1}}M_{a_{2k+1}a},\quad(k=0,1,2,\ldots).
\end{eqnarray}
The algebra (\ref{2.9}) can be considered as another extension with the help
of the vector generator ${1\over2}(P+D)_a$ and tensor generator $Z_{ab}$ of
the Poincar\'e algebra formed by the generators $M_{ab}$ and
${1\over2}(P-D)_a$. By using the expressions (\ref{2.7}) and (\ref{2.8}), the
Casimir operators (\ref{2.10}) can be represented in the form
\begin{eqnarray}
cZ^{aa_1}Z_{a_1a_2}\cdots
Z_{a_{2k-1}a_{2k}}Z^{a_{2k}a_{2k+1}}S_{a_{2k+1}a},\quad(k=0,1,2,\ldots).\nonumber
\end{eqnarray}

\section{Supersymmetric generalization}

The extended Poincar\'e algebra (\ref{2.2}) admits the following supersymmetric
generalization:
\begin{eqnarray}
\{Q_\alpha,Q_\beta\}=-d(\sigma^{ab}C)_{\alpha\beta}Z_{ab},\nonumber
\end{eqnarray}
\begin{eqnarray}
[M_{ab},Q_\alpha]=-(\sigma_{ab}Q)_\alpha,\nonumber
\end{eqnarray}
\begin{eqnarray}
[P_a,Q_\alpha]=0,\nonumber
\end{eqnarray}
\begin{eqnarray}\label{3.1}
[Z_{ab},Q_\alpha]=0
\end{eqnarray}
with the help of the super-translation generators
\begin{eqnarray}
Q_\alpha=-\left[{\partial\over\partial {\bar\theta}^{\alpha}}+
{d\over2}(\sigma^{ab}\theta)_\alpha{\partial\over\partial z^{ab}}\right],\nonumber
\end{eqnarray}
where $\theta=C\bar\theta$ is a Majorana Grassmann spinor, $C$ is a charge conjugation
matrix, $d$ is some constant and $\sigma_{ab}={1\over4}[\gamma_a,\gamma_b]$.

The rotation generators acquire the terms depending on the Grassmann variables
$\theta_\alpha$
\begin{eqnarray}
M_{ab}=x_a{\partial\over\partial x^b}-x_b{\partial\over\partial x^a}
+{z_a}^c{\partial\over\partial z^{bc}}-{z_b}^c{\partial\over\partial z^{ac}}
-(\sigma_{ab}\theta)_\alpha{\partial\over\partial\theta_\alpha}+S_{ab},\nonumber
\end{eqnarray}
whereas the expressions (\ref{2.7}) for the translations $P_a$ and tensor
generator $Z_{ab}$ remain unchanged.

The validity of the Jacobi identities
\begin{eqnarray}
[P_a,\{Q_\alpha,Q_\beta\}]=\{Q_\alpha,[P_a,Q_\beta]\}+
\{Q_\beta,[P_a,Q_\alpha]\}\nonumber
\end{eqnarray}
and
\begin{eqnarray}
[M_{ab},\{Q_\alpha,Q_\beta\}]=\{Q_\alpha,[M_{ab},Q_\beta]\}+
\{Q_\beta,[M_{ab},Q_\alpha]\}\nonumber
\end{eqnarray}
for the supersymmetric generalization of the extended Poincar\'e algebra
(\ref{2.2}) verified for the dimensions $D=2,3,4$ with the use of the symmetry
properties of the matrices $C$ and $\gamma_aC$ and the relations
(\ref{A.1})--(\ref{A.3}) of the Appendix.

One of the generating Casimir operator in the dimensions $D=2,3$ is generalized
 into the following form:
\begin{eqnarray}\label{3.2}
P^aP_a+cZ^{ab}M_{ba}-{c\over2d}Q_\alpha(C^{-1})^{\alpha\beta}Q_\beta,
\end{eqnarray}
while the form of the rest generating Casimir operators in these dimensions
are not changed.
Note that in the case $D=3$ there is also the following Casimir operator:
\begin{eqnarray}
Z^aQ_\alpha(C^{-1}\gamma_a)^{\alpha\beta}Q_\beta.\nonumber
\end{eqnarray}

One of the simplest Casimir operator (\ref{2.6}) in $D=4$
is also generalized into the form (\ref{3.2}). The supersymmetric
generalization of the
more complicated Casimir operators in the four-dimensional case has the
following structure:
\begin{eqnarray}
P^aZ_{ab}Z^{bc}P_c+cZ^{ab}Z_{bc}Z^{cd}M_{da}
+{2c\over5d}Q_\alpha(C^{-1}\sigma^{ab}Z_{ab}\sigma^{cd}Z_{cd})^{\alpha\beta}Q_\beta
+{c\over2d}Z^{ab}Z_{ab}Q_\alpha(C^{-1})^{\alpha\beta}Q_\beta,\nonumber
\end{eqnarray}
\begin{eqnarray}
P^aZ_{ab}Z^{bc}Z_{cd}Z^{de}P_e&+&cZ^{ab}Z_{bc}Z^{cd}Z_{de}Z^{ef}M_{fa}\cr\nonumber\\
&+&{2c\over5d}Q_\alpha\left[C^{-1}\sigma^{ab}Z_{ab}\sigma^{cd}\left(Z_{ce}Z^{ef}Z_{fd}
+{3\over10}Z^{gh}Z_{hg}Z_{cd}\right)\right]^{\alpha\beta}Q_\beta\cr\nonumber\\
&-&{c\over20d}\left[7Z_{ab}Z^{bc}Z_{cd}Z^{da}
+3(Z^{ef}Z_{fe})^2\right]Q_\alpha(C^{-1})^{\alpha\beta}Q_\beta.\nonumber
\end{eqnarray}
An algorithm for the construction of the supersymmetric generalization of the
Casimir operators (\ref{2.4}) is obvious and based on the use of the following
commutation relations:
\begin{eqnarray}
\left[{1\over2d}Q_\alpha(C^{-1})^{\alpha\beta}Q_\beta,Q_\gamma\right]=Z^{ab}(\sigma_{ab}Q)_\gamma,\nonumber
\end{eqnarray}
\begin{eqnarray}
\left[{2\over5d}Q_\alpha(C^{-1}\sigma^{ab}Z_{ab}\sigma^{cd}\tilde Z_{cd})^{\alpha\beta}Q_\beta,
Q_\gamma\right]=\biggl(Z^{ab}Z_{bc}\tilde Z^{cd}
&+&{7\over10}Z_{bc}\tilde Z^{cb}Z^{ad}\cr\nonumber\\ 
&+&{3\over10}Z_{bc}Z^{cb}\tilde Z^{ad}\biggr)(\sigma_{ad}Q)_\gamma,\nonumber
\end{eqnarray}
where
\begin{eqnarray}
\tilde Z^{ab}=Z^{aa_1}Z_{a_1a_2}\cdots Z_{a_{2k-1}a_{2k}}Z^{a_{2k}b},\quad
(k=0,1,\ldots).\nonumber
\end{eqnarray}

\section{Conclusion}

Thus, in the present report we proposed the extension of the Poincar\'e
algebra with the help of the second rank tensor generator. Casimir operators
for the extended algebra are constructed. The form of the Casimir operators
indicate that an irreducible representation of the extended algebra contains
the fields with the different masses.
A consideration is performed for the arbitrary dimensions $D$.
A possible supersymmetric generalization
of the extended Poincar\'e algebra is also given for the
particular cases with the dimensions $D=2,3,4$.

It would be interesting to find the spectra of the Casimir operators and to
construct the models based on the extended Poincar\'e algebra. The work in
these directions is in progress.

\appendix
\section{Appendix}

As a real (Majorana) representation for the two-dimensional $\gamma$-matrices 
and charge conjugation matrix $C$ we adopt
\begin{eqnarray}
\gamma^0=C=-C^T=-i\sigma_2,\quad\gamma^1=\sigma_1,\quad\gamma_5=
{1\over2}\epsilon^{ab}\gamma_a\gamma_b=\sigma_3;\nonumber
\end{eqnarray}
\begin{eqnarray}
\{\gamma_a,\gamma_b\}=2g_{ab},\quad g_{11}=-g_{00}=1,
\quad C^{-1}\gamma_aC=-{\gamma_a}^T,\nonumber
\end{eqnarray}
where $\sigma_i$ are Pauli matrices. The matrices $\gamma_a$ satisfy the 
relations
\begin{eqnarray}
\gamma_a\gamma_5=\epsilon_{ab}\gamma^b,\quad\gamma_a\gamma_b=g_{ab}-
\epsilon_{ab}\gamma_5. \label{A.1}
\end{eqnarray}

For the Majorana three-dimensional $\gamma$-matrices and charge conjugation 
matrix  $C$ we take
\begin{eqnarray}
\gamma^0=C=-C^T=-i\sigma_2,\quad\gamma^1=\sigma_1,\quad\gamma^2=\sigma_3;
\nonumber
\end{eqnarray}
\begin{eqnarray}
\{\gamma_a,\gamma_b\}=2g_{ab},\quad g_{11}=g_{22}=-g_{00}=1,\quad
C^{-1}\gamma_aC=-{\gamma_a}^T.
\nonumber
\end{eqnarray}
The matrices $\gamma_a$ obey the relations
\begin{eqnarray}
\gamma_a\gamma_b=g_{ab}-\epsilon_{abc}\gamma^c. \label{A.2}
\end{eqnarray}

At last, the real four-dimensional $\gamma$-matrices and matrix $C$ are
\begin{eqnarray}
\gamma^0=C=-C^T=-i\left(\begin{array}{cc}0&\sigma_2\\
\sigma_2&0\end{array}\right),\quad
\gamma^1=\left(\begin{array}{cc}\sigma_3&0\\
0&\sigma_3\end{array}\right),\nonumber
\end{eqnarray}
\begin{eqnarray}
\gamma^2=i\left(\begin{array}{cc}0&\sigma_2\\
-\sigma_2&0\end{array}\right),\quad
\gamma^3=-\left(\begin{array}{cc}\sigma_1&0\\
0&\sigma_1\end{array}\right),
\nonumber
\end{eqnarray}
\begin{eqnarray}
\{\gamma_a,\gamma_b\}=2g_{ab},\quad g_{11}=g_{22}=g_{33}=-g_{00}=1,\quad
C^{-1}\gamma_aC=-{\gamma_a}^T,\quad\gamma_5=
{1\over4}\epsilon^{abcd}\gamma_a\gamma_b\gamma_c\gamma_d.\nonumber
\end{eqnarray}
The matrices $\gamma_a$ and $\sigma_{ab}$ meet the relations
\begin{eqnarray}
\gamma_a\sigma_{bc}={1\over2}\epsilon_{abcd}\gamma^d\gamma_5+
{1\over2}(\gamma_cg_{ab}-\gamma_bg_{ac}),\quad
\sigma_{ab}\gamma_c={1\over2}\epsilon_{abcd}\gamma^d\gamma_5+
{1\over2}(\gamma_ag_{bc}-\gamma_bg_{ac}),\nonumber
\end{eqnarray}
\begin{eqnarray} 
\sigma_{ab}\sigma_{cd}={1\over4}(g_{ad}g_{bc}-g_{ac}g_{bd}-
\epsilon_{abcd}\gamma_5)
+{1\over2}(\sigma_{ad}g_{bc}+\sigma_{bc}g_{ad}-\sigma_{ac}g_{bd}-
\sigma_{bd}g_{ac}).\label{A.3}
\end{eqnarray}

\subsection*{Acknowledgments}

One of the authors (V.A.S.) would like to thank B.A. Dubrovin for the useful
discussions. V.A.S. is sincerely grateful to L. Bonora for the fruitful
discussions and kind hospitality at SISSA/ISAS (Trieste), where the main part
of this work has been performed.

\LastPageEnding

\end{document}